\documentclass[11pt,twoside]{article} 
\begin{document}
\def\ga{\alpha}
\def\gb{\beta}
\def\gD{\Delta}
\def\gG{\Gamma}
\def\ge{\epsilon}
\def\gg{\gamma}
\def\gd{\delta}
\def\gf{\phi}
\def\gvf{\varphi}
\def\phid{\phi^\dagger}
\def\gm{\mu}
\def\gn{\nu}
\def\gO{\Omega}
\def\gp{\pi}
\def\gP{\Pi}
\def\gr{\rho}
\def\gs{\sigma}
\def\gS{\Sigma}
\def\gch{\chi}
\def\gCh{\Chi}
\def\gl{\lambda}
\def\gL{\Lambda}
\def\gth{\theta}
\def\gtt{\theta}
\def\go{\omega}
\def\gz{\zeta}
\def\ad{a^\dagger}
\def\dg{\dagger}
\def\delp{\partial_+}
\def\delm{\partial_-}
\def\delpm{\partial_\pm}
\def\delmu{\partial_\gm}
\def\delmuu{\partial^\gm}
\def\delnu{\partial_\gn}
\def\xpm{x^{\pm}}
\def\delrlp{\stackrel {\leftrightarrow} {\partial_+}}
\def\delrlm{\stackrel {\leftrightarrow} {\partial_-}} 
\def\delrlmu{\stackrel {\leftrightarrow} {\partial_\gm}}
\def\delrlz{\stackrel {\leftrightarrow} {\partial_0}}
\def\delrl1{\stackrel {\leftrightarrow} {\partial_1}}
\def\delrli{\stackrel {\leftrightarrow} {\partial_i}}
\def\delrlmup{\stackrel {\leftrightarrow} {\partial^\mu}}
\def\tdphi{\stackrel{\dot} {\varphi}}
\def\vecA{\stackrel{\rightarrow} {A}}
\def\vecp{\stackrel{\rightarrow} {p}}
\def\vecx{\stackrel{\rightarrow} {x}}
\def\part{\partial}
\def\parti{\partial_i}
\def\hlf{\frac{1}{2}}
\def\A0{A^{+}_0}
\def\psip{\psi_+}
\def\psin{\psi_-}
\def\psih{\psi_2}
\def\psihd{{\psi_2}^\dagger}
\def\psil{\psi_1}
\def\psild{{\psi_1}^\dagger}
\def\phih{\phi_2}
\def\phihd{{\phi_2}^\dagger}
\def\phil{\phi_1}
\def\phild{{\phi_1}^\dagger}
\def\psipd{\psi^{\dagger}_+}
\def\psind{\psi^{\dagger}_-}
\def\Psih{\Psi_2}
\def\Psihd{{\Psi_2}^\dagger}
\def\Psil{\Psi_1}
\def\Psild{{\Psi_1}^\dagger}
\def\psib{\overline{\psi}}
\def\chib{\overline{\chi}}
\def\phib{\overline{\phi}}
\def\phid{\phi^\dagger}
\def\Phib{\overline{\Phi}}
\def\Phid{\Phi^\dagger}
\def\Psib{\overline{\Psi}}
\def\Psid{\Psi^\dagger}
\def\ub{\overline{u}(p)}
\def\vb{\overline{v}(p)}
\def\psid{\psi^{\dag}}
\def\chid{\chi^{\dag}}
\def\sla#1{#1\!\!\!/}
\def\xpl{x^{+}}
\def\ypl{y^{+}}
\def\zpl{z^{+}}
\def\xmin{x^{-}}
\def\ymin{y^{-}}
\def\zmin{z^{-}}
\def\xp{x_\perp}
\def\yp{y_\perp}
\def\pp{p_\perp}
\def\cubr{\big\{}
\def\cubl{\big\}}
\def\ulix{\underline{x}}
\def\uliy{\underline{y}}
\def\ulip{\underline{p}}
\def\ulip{\underline{p}}
\def\ulin{\underline{n}}
\newcommand{\nc}{\newcommand}
\nc{\intl}{\int\limits_{-L}^{+L}\!\frac{{\rm d}x^-}{2}}
\nc{\intly}{\int\limits_{-L}^{+L}\!{{{\rm d}y^-}\over\!2}}
\nc{\intlz}{\int\limits_{-L}^{+L}\!{{{\rm d}z^-}\over\!2}}
\nc{\intlu}{\int\limits_{-L}^{+L}\!{{{\rm d}u^-}\over\!2}}
\nc{\intlv}{\int\limits_{-L}^{+L}\!{{{\rm d}v^-}\over\!2}}
\nc{\intv}{\int\limits_{-V}^{}\!{\rm d}^3\ulix}
\nc{\intvy}{\int\limits_{-V}^{}\!{\rm d}^3\uliy}
\nc{\zmint}{\int\limits_{-L}^{+L}\!{{{\rm d}x^-}\over{\!2L}}}
\nc{\zminty}{\int\limits_{-L}^{+L}\!{{{\rm d}y^-}\over{\!2L}}}
\nc{\intp}{\int\limits_{0}^{+\infty}\!{{{\rm d}p^+}\over\!4\gp}}
\nc{\inp}{\int\limits_{0}^{\infty}\!{{{\rm d}p^+}\over{\!2\sqrt{2\gp}}}}
\nc{\inq}{\int\limits_{0}^{\infty}\!{{{\rm d}q^+}\over{\!2\sqrt{2\gp}}}}
\nc{\inpp}{\int\limits_{0}^{\infty}\!{{{\rm d}p^+}\over{\!2p^+\sqrt{2\gp}}}}
\nc{\inqq}{\int\limits_{0}^{\infty}\!{{{\rm d}q^+}\over{\!2q^+\sqrt{2\gp}}}}
\nc{\insl}{\int\limits_{-L}^{+L}\!{\rm d}x}
\nc{\intex}{\int\limits_{-\infty}^{+\infty}\!{\rm d}x^1}
\nc{\intep}{\int\limits_{-\infty}^{+\infty}\!{\rm d}p^1}
\nc{\inteq}{\int\limits_{-\infty}^{+\infty}\!{\rm d}q^1}
\nc{\intek}{\int\limits_{-\infty}^{+\infty}\!{\rm d}k^1}
\nc{\intel}{\int\limits_{-\infty}^{+\infty}\!{\rm d}l^1}
\nc{\inty}{\int\limits_{-\infty}^{+\infty}\!{\rm d}y^-}
\nc{\intz}{\int\limits_{-\infty}^{+\infty}\!{\rm d}z^-}
\def\beq{\begin{equation}}
\def\eeq{\end{equation}}
\def\bea{\begin{eqnarray}}
\def\eea{\end{eqnarray}}
\nc{\intx}{\int\limits_{-\infty}^{+\infty}\!\frac{{\rm d}x^-}{2}}
\nc{\intgix}{\int\limits_{-\infty}^{+\infty}\!{{{\rm d}x^-}\over\!2}}
\nc{\intgiy}{\int\limits_{-\infty}^{\infty}\!{{{\rm d}y^-}\over\!2}}
\def\Ra{\Rightarrow}
\def\tc{\textcolor}
\def\trip{\stackrel{.} {:}}


\title{\bf HAMILTONIANS AND PHYSICAL VACUA \\OF EXACTLY SOLVABLE MODELS\thanks
{Talk given at the HSQCD 2010 conference, Gatchina, Russia, July 5-9, 
2010}}
\author{\sf{L$\!\!$'ubom\'{\i}r MARTINOVI\u{C}}\\
{\it{BLTP JINR, 141 980 Dubna, Russia}}\\ 
{\it and}  
\\
{\it{Institute of Physics, Slovak Academy of Sciences}}\\
{\it{D\'ubravsk\'a cesta 9, 845 11 Bratislava, Slovakia}}\\
{\it{E-mail}}: fyziluma@savba.sk}
\maketitle
\begin{abstract} 
\noindent
Correct quantum Hamiltonians of a few exactly solvable models in two 
space-time dimensions are derived by taking into account 
operator solutions of the field equations. While two versions of the  
model with derivative-coupling are found to be equivalent in many respects to a 
free theory, physical vacua of the massless  
Thirring and Federbush models are obtained by means of a Bogoliubov 
transformation in the form of a coherent state quadratic in composite boson 
operators. Contrary to the conventional treatment, the Federbush model is   
shown to have the same interacting structure in both space-like and light-front 
formulations. 
\end{abstract}  



\section{Introduction}

Exactly solvable models, i.e. simple relativistic theories in D=1+1 in which  
operator solutions of field equations are known, provide us with a suitable 
arena for analyzing the nonperturbative structure of the usual  
("space-like" - SL) and light-front (LF) forms \cite{Dir49} of hamiltonian 
field theory and also for a comparison between them. In this paper, we will 
study models with derivative coupling (DCM) \cite{Sch61,RS,Belve},  
the massless Thirring model (TM) \cite{Thir} and the massive Federbush model 
(FM) \cite{Federb}. The main new idea is to make use of the knowledge of the 
operator solutions to express the Lagrangian and Hamiltonian entirely in terms 
of true field degrees of freedom, which are actually free fields. In the case 
of the TM and FM, the diagonalization of the SL Hamiltonian  
by means of a Bogoliubov transformation generates the 
true vacuum, i.e. the lowest-energy eigenstate of the full Hamiltonian, not 
just of its free part. This vacuum state is a transformed Fock vacuum - a  
coherent state quadratic in composite boson operators that represent currents.   

\section{Derivative-coupling models}
The general Lagrangian density of the model with derivative coupling is  
\beq
{\cal L}={\cal L}_0+{\cal L^{'}},~~ {\cal L}_0=\frac{i}{2}\Psib\gg^\mu\delrlmu 
\Psi - m\Psib\Psi + 
\hlf \delmu \phi 
\delmuu \phi - \hlf \mu^2 \phi^2,~~ {\cal L{'}}= - g\delmu \phi K^\mu. 
\label{DCML}
\eeq
$K^\mu$ can be the vector current $J^\mu(x)$ or the axial-vector current 
$J_5^\mu(x)$. 
We will briefly analyze the case with $J^\mu_5$ and  
$m=0$ (the Rothe-Stamatescu (RS) model \cite{RS}). The same model with 
massive fermions is not exactly solvable \cite{Belve}. 
The corresponding field equations   
\bea 
i\gg^\mu\delmu \Psi = g\delmu \phi\gg^\mu \gg^5\Psi,~~~ 
\delmu \delmuu \phi + \mu^2 \phi = g\delmu J_5^\mu = 0
\label{DCME}
\eea
are solved in terms of the free scalar field $\phi(x)$ and the 
massless fermion field $\psi(x)$ as 
\bea 
&&\!\!\!\!\!\!\!\!\!\!\!\Psi(x) = :e^{-ig\gg^5\phi(x)}:\psi(x),
~~\psi(x) = \int\limits^{+\infty}_{-\infty} 
\frac{dp^1}{\sqrt{2\gp}} 
\big\{b(p^1)u(p^1)e^{-ip.x} + 
d^\dagger(p^1)v(p^1)e^{ip.x}\big\}, \label{solut} \\ 
&&\!\!\!\!\!\!\!\!\!\!\!\phi(x) = \frac{1}{\sqrt{2\gp}}\int\limits^{+\infty}_
{-\infty}
\frac{dk^1}{\sqrt{2E(k^1)}}\big[a(k^1)e^{-i\hat{k}.x} + 
a^\dagger(k^1)
e^{i\hat{k}.x}\big],  
~\hat{k}.x \equiv E(k^1)t - k^1x^1,  
\label{DCMS}
\eea 
$E(k^1)=\sqrt{k_1^2 + \mu^2}$.   
Since the solution (\ref{DCMS}) tells us that the interacting fermion field 
$\Psi(x)$ is composed from two free fields, we should formulate the  
dynamics of the model in terms of these degrees of freedom by 
inserting the solution to the Lagrangian. 
Consequently, the contribution of 
the derivative term in ${\cal L}_0$ eliminates the interaction term! Thus\\
$H = H_{0} = 
\intex \big[-i\psi^\dagger \ga^1\partial_1\psi + 
\hlf(\partial_0 \phi)^2 + \hlf(\partial_1 \phi)^2 + \hlf
\mu^2\phi^2\Big]$.  
Clearly, $\vert vac \rangle \equiv \vert 0 \rangle$.  
The signature of an interacting theory is  
$g$-dependent correlation functions calculated from (\ref{solut}). 
With $D^{(+)}(x-y) = 
\langle 0 \vert \phi(x)\phi(y)\vert 0 \rangle$, $S^{(+)}(x-y) = 
\langle 0\vert \psi(x)
\psib(y)\vert 0 \rangle$, we get e.g.  
\beq
\langle 0 \vert \Psi(x)\Psib(y)\vert 0 \rangle = e^{g^2 D^{(+)}(x-y)}
S^{(+)}(x-y).  
\label{corf}
\eeq
Note that conventionally one replaces $\Psi(x)$ by the free field in 
${\cal L}_0$. 
This procedure yields 
\beq 
H^{'} = \frac{g}{2}\intek \frac{k^1 \vert k^1\vert^{1/2}}
{\sqrt{\gp E(k^1)}} \big[a^\dagger(k^1)c^\dagger(-k^1) + a(k^1)c(-k^1) -  
a^\dagger(k^1)c(k^1) - c^\dagger(k^1)a(k^1)\big],   
\label{hint}
\eeq
i.e. a nondiagonal interaction Hamiltonian expressed in terms of 
\bea
&&c(k^1) = \frac{i}{\sqrt{k^0}}\int dp^1\big\{\theta\big(p^1k^1)\big)
\big[b^\dagger(p^1)b(p^1+k^1) - d^\dagger(p^1)d(p^1+k^1)\big] + \nonumber \\
&&~~~~~~~~~  + \ge(p^1)\theta\big(p^1(p^1-k^1)\big)d(k^1-p^1)b(p^1)\big\},~~
~~[c(k^1),c^\dagger(l^1)]=\gd(k^1-l^1),  
\eea
that define the bosonized current 
\cite{Klaib} $j^\mu(x) = -\frac{i}{\sqrt{2}\pi}\int\frac{dk^1}
{\sqrt{2k^0}}k^\mu
\big\{
c(k^1)e^{-i\hat{k}.x} - c^\dagger(k^1)e^{i\hat{k}.x}\big\}$. 
Diagonalization of the full Hamiltonian by a suitable unitary operator 
$U = \exp(iS)$ would then generate also the new vacuum  
$\vert \gO \rangle = N \exp\big[\gg(g)\intek c^\dagger(-k^1)
a^\dagger(k^1)\big]\vert 0 \rangle$.
The two methods lead to a completely different vacuum 
structure of the RS model. The second (standard) treatment is incorrect. 
The same conclusions are valid also for the massive model 
with $J^\mu$ \cite{lmvalc}, where moreover the SL results  
fully agree with a paralell LF analysis. 

\section{The Thirring model}
The operator solution of the Thirring model was given by Klaiber \cite{Klaib} 
who also calculated n-point correlation functions.  
Despite of having been studied intensively in the past, some of its 
properties have not been fully understood.   
Here we sketch a novel systematic Hamiltonian study based on the model's 
solvability.  
We start from the Lagrangian density 
\beq
{\cal L}=\frac{i}{2}\Psib\gg^\mu\delrlmu \Psi - \hlf gJ_\mu J^\mu,~~
J^\mu=\Psib \gg^\mu \Psi,~~ 
\partial_\mu J^\mu(x) = 0.
\label{TL}
\eeq
The field equations are  
$i\gg^\mu\partial_\mu \Psi(x) = gJ^\mu(x)\gg_\mu\Psi(x)$.  
The simplest solution is 
\beq
\Psi(x) = :e^{-i(g/\sqrt{\pi})j(x)}:\psi(x),
~~\gg^\mu\partial_\mu\Psi(x) = 0,~ 
j_\mu(x) = \frac{1}{\sqrt{\pi}}\partial_\mu j(x),~~J^\mu(x) = j^\mu(x).
\label{siso}
\eeq
$j(x)$ is the "integrated current". 
Free fields define the solution of the interacting model.
The correct Hamitonian is obtained by inserting the solution (\ref{siso}) 
to the Lagrangian. This reverses the sign of the interaction term 
also in the canonically obtained Hamiltonian 
$H = H_0 + H_{g} = \intex \Big[-i\psi^\dagger\ga^1\partial_1 \psi -   
\frac{1}{2}g\big(j^0j^0 - j^1j^1\big)\Big]$ (conventionally, one has $+\hlf 
g$),   
\beq  
H = \intek \vert k^1 \vert \Big\{b^\dagger(k^1)b(k^1) + 
d^\dagger(k^1)d(k^1) + 
\frac{g}{\gp} \Big[c^\dagger(k^1)
c^\dagger(-k^1)+ c(k^1)c(-k^1)\Big]\Big\}. 
\label{TMH}
\eeq 
To diagonalize $H_{g}$, we define the operator $T$ with the same commutation  
property as $H_0$, 
$T = \intek \vert k^1 \vert c^\dagger(k^1)c(k^1)$,~ 
$\big[T,c(k^1)\big] = -\vert k^1 \vert c(k^1)$,
and the unitary operator 
$U = e^{iS}$,
\beq
S = -\frac{i}{2}\intep \gg(p^1)\big[c^\dagger(p^1)
c^\dagger(-p^1) 
- c(p^1)c(-p^1)\big].
\label{S}
\eeq
We form new Hamiltonians
$\hat{H}_0 = H_0 - T$, $\hat{H}_{g} = H_{g} + T$.
Due to $\big[S,\hat{H}_0\big] = 0$, $\hat{H}_0$ is invariant 
with respect to $U$, 
while $\hat{H}_{g}$ transforms non-trivially, since    
$\big[S,c(k^1)\big] = i\gg(k^1)c^\dagger(-k^1)$, 
implying $c(k^1) \rightarrow e^{iS}c(k^1)e^{-iS} = c(k^1)\cosh \gg(k^1) - 
c^\dagger(-k^1) 
\sinh \gg(k^1)$. Consequently the interacting Hamiltonian acquires the most 
general operator form. It becomes diagonal,
\beq
\hat{H}_{int} = ({\rm cosh}2\gg_d)^{-1}\intek \vert k^1 \vert c^\dagger(k^1)
c(k^1),
\label{diah}
\eeq
if $\gg(k^1) = \gg_d = \hlf {\rm arctanh} \frac{g}{\gp}$.
Thus, we have achieved
$e^{iS}\hat{H}_{g}e^{-iS}\vert 0 \rangle = 0$. This implies that the state
$\vert \gO \rangle = e^{-iS}\vert 0 \rangle$ is the true ground state of the 
original Hamiltonian. 
Explicitly, 
\beq
\vert \gO \rangle = \exp\Big[-\frac{1}{2}\gg_d\intep \big[c^\dagger(p^1)
c^\dagger(-p^1) - c(p)c(-p^1)\big]\Big]\vert 0 \rangle.  
\label{ftv}
\eeq
It is a coherent state of effective bosons, bilinear in fermion Fock 
operators. A direct computation shows that this state has zero momentum and 
axial charge. There is no spontaneous symmetry breaking, contrary to some 
statements in literature based on NJL type of approximations.    
The correlation functions, like for example 
$C_2(x-y) = \langle vac \vert \Psi(x)\Psib(y)\vert vac \rangle$,   
should now be calculated from the normal-ordered  
operator solution (\ref{siso}) using an infrared cutoff and the vacuum 
state $\vert \gO \rangle$. In the conventional treatment, one uses  
the Fock vacuum $\vert 0 \rangle$. 
Performing all necessary commutations, one finds  
\bea 
&&\!\!\!\!\!\!\!\!\!\!\!C_2(x-y)=e^{\frac{g^2}{\gp}D^{(+)}(x-y)}
e^{-2g\big[D^{(+)}(x-y)) 
+ \gg^5 \tilde{D}^{(+)}(x-y)\big]}
\langle 0 \vert \psi(x)\psib(y)\vert 0 \rangle, 
\nonumber \\  
&&\!\!\!\!\!\!\!\!\!\!\!D^{(+)}(x) = \frac{1}{2\gp}\int\limits_{-\infty}^
{+\infty} 
\frac{dk^1}
{2\vert k^1 \vert}\theta(\vert k^1 
\vert - \gl)e^{-ik.x} = -\frac{1}{4\gp}
\ln\big(-\mu^2 x^2 + ix^0\ge\big),~  
\mu = e^{\gg_E}\gl.  
\label{Dp}
\eea 
$\tilde{D}^{+}(x)$ has $\ge(k^1)$ in the integrand.  
We then find   
$\langle \gO \vert \Psi(x)\Psib(y)\vert \gO \rangle = F_2(x-y;\kappa)
C_2(x-y)$. $F_2(x-y;\kappa(g))$ is a complicated function equal to unity for  
$\kappa(g) = 0$, i.e. when $\vert \gO\rangle \rightarrow \vert 0 \rangle$.   

\section{The Federbush model}
The Federbush model (FM) \cite{Federb} is the only known {\it massive} 
solvable model. Its true physical SL ground state can be found analogously  
to the massless Thirring model. This task requires a
generalization of the Klaiber's bosonization to massive fermions. The model 
offers us a unique opportunity to solve a field theory nonperturbatively 
in both SL and LF forms and to compare their structures. 
The Lagrangian of the FM 
describes two species of the fermion field interacting via specific current-- 
current coupling, 
\bea 
{\cal L}=\frac{i}{2}\Psib\gg^\mu\delrlmu \Psi - m\Psib\Psi +  
 \frac{i}{2}\Phib\gg^\mu\delrlmu \Phi - \mu\Phib\Phi - g\ge_{\mu\nu}J^\mu
H^\nu.  
\label{FL}
\eea 
The currents are $J^\mu =\Psib\gg^\mu\Psi,~H^\mu=\Phib\gg^\mu\Phi$. 
The coupled field equations read  
\beq 
i\gg^\mu\delmu\Psi(x) = m\Psi(x) + g\ge_{\mu\nu}\gg^\mu H^\nu(x)\Psi(x),~  
i\gg^\mu\delmu\Phi(x) = \mu\Phi(x) - g\ge_{\mu\nu}\gg^\mu J^\nu(x)\Phi(x).  
\label{Fel}
\eeq 
The relations $J^\mu(x) = \ge^{\mu\nu}\delnu j(x)/\sqrt{\gp},~ 
H^\mu(x) = \ge^{\mu\nu}\delnu h(x)/\sqrt{\gp}$ define the "integrated 
currents" $j(x)$ and $h(x)$.  
The latter enter into the solutions in an "off-diagonal" way: 
\beq
\Psi(x) = e^{-i\frac{g}{\sqrt{\gp}}h(x)}\psi(x),~~~~
\Phi(x) = e^{i\frac{g}{\sqrt{\gp}}j(x)}\phi(x),~  
\label{Fsol}
\eeq
Here the free fields $\psi(x)$ and $\phi(x)$ are defined by 
$i\gg^\mu\delmu\psi(x) = m\psi(x)$,~
$i\gg^\mu\delmu\phi(x) = \mu\phi(x)$. The above solutions also imply 
$J^\mu(x)=j^\mu(x)$, $H^\mu(x)=h^\mu(x)$. 
Exponentials of the composite fields are more singular than in the 
massless case and have to be defined using the "triple-dot ordering" 
\cite{Wig}.
We avoid this by bosonization of the massive current. 
 
The usual treatment yields contradictory picture of the dynamics of the 
model - the SL Hamiltonian contains interaction while the LF one has  
the free form:   
\bea  
&&H=\intex \Big[-\frac{i}{2}\psid\ga^1\delrl1 \psi + m\psid\gg^0\psi -  
 \frac{i}{2}\phid\ga^1\delrl1\phi + \mu\phid\gg^0\phi -   
gj^0h^1 + gj^1h^0\Big], \nonumber \\ 
&&P^- = \intx\Big[m\Big(\psild\psih + \psihd\psil\big) + \mu\Big(\phild\phih + 
\phihd\phil\Big)\Big].
\label{Fslfha}
\eea 
Our approach leads to a different picture. Inserting  
(\ref{Fsol}) into the Lagrangian, we get  
\bea  
&&{\cal L} = \frac{i}{2}\psid\gg^0 \gg^\mu\delrlmu \psi - m\psib\psi +  
 \frac{i}{2}\phid\gg^0 \gg^\mu\delrlmu\phi - \mu\phib\phi + g\ge_{\mu\nu}j^\mu
h^\nu, \nonumber \\  
&&H = \intex \big[-i\psid\ga^1\partial_1\psi + m\psib\psi -i\phid\ga^1
\partial_1\phi + \mu\phib\phi + g\big(j^0h^1 - j^1h^0\big)\big]. 
\label{LH}
\eea 
Both operators have an 
opposite sign (with respect to the conventional result) in the interaction 
piece. 
The interaction term is non-diagonal when expressed in terms of bosonized 
massive currents. A Bogoliubov transformation is required to diagonalize it.  
The massive analogues of Klaiber's operators $c(k^1)$ are 
surprisingly complicated \cite{LP2}. 

The LF form of the Lagrangian is     
\bea
&&{\cal L}_{lf} = i\Psi^\dagger_2\delrlp\Psi_2 + 
i\Psi_1^\dagger\delrlm\Psi_1 - 
m\big(\Psi_2^\dagger\Psi_1 + 
\Psi_1^\dagger\Psi_2\big) + 
i\Phi^\dagger_2\delrlp\Phi_2 +  
i\Phi_1^\dagger\delrlm\Phi_1 - \nonumber \\ 
&&~~~~~~~~~~~~~~~~~~~~~~- \mu\big(\Phi_2^\dagger\Phi_1 + 
\Phi_1^\dagger\Phi_2\big) - 
\frac{g}{2}j^+h^- + 
\frac{g}{2}j^-h^+.  
\label{Fedlfl}
\eea
The field equations are   
\bea
&&\!\!\!\!\!\!2i\delp \Psi_2(x) = m\Psi_1 - g h^-\Psi_2,~~2i \delm \Psi_1 = 
m\Psi_2 + 
gh^+\Psi_1, \nonumber \\
&&\!\!\!\!\!\!2i\delp \Phi_2(x) = \mu\Phi_1 + g j^-\Phi_2,~~~~
2i \delm \Phi_1 = 
\mu\Phi_2 - 
gj^+\Phi_1.
\label{seqs}
\eea
The currents are  
$j^+(x) = 2:\psi_2^\dagger(x)\psi_2(x):$, $j^-(x) =  
2:\psi_1^\dagger(x)\psi_1(x):$,  
$h^+(x) = 2:\phi_2^\dagger(x)\phi_2(x):$, $h^-(x) =  
2:\phi_1^\dagger(x)\phi_1(x)$. 
Equations (\ref{seqs}) are solved by (\ref{Fsol}) in terms of the free LF 
fields and integrated currents.   
The bosonized form of the LF Hamiltonian is quadratic  
and diagonal:
\beq 
P^-_g = \frac{g}{8\gp}\intek k^+ \Big\{\big[A^\dg(k^+)D(k^+) 
- B^\dg(k^+)C(k^+)\big] + H.c.\Big\}
\label{boslfh}
\eeq 
The operators $A(k^+),~B(k^+),~C(k^+)$ and $D(k^+)$ correspond to $j^+,
~h^+,~j^-$ and $h^-$. Their form is as simple as the massless $c(k^1)$ in  
the SL case.   
Complexities will enter in calculations of the correlation functions since the 
composite LF boson operators do not commute to the delta function at unequal 
LF times \cite{LP2}. 
It will be    
interesting to analyze how the SL and LF schemes generate mutually consistent 
results for the correlators given the different vacuum 
structure in the two hamiltonian forms of the 
Federbush model. 
\section*{Acknowledgements}
This work was done in collaboration with P. Grang\'e. It was supported by the 
VEGA grant No. 2/0070/2009 and by IN2P3 
funding at the Universit\'e Montpellier II.  

\end{document}